\documentclass[10pt,journal]{IEEEtran}
\usepackage{cite}
\usepackage{amsmath,amssymb,amsfonts}
\usepackage{algorithmic}
\usepackage{algorithm}
\usepackage{graphicx}
\usepackage{textcomp}
\usepackage{makecell}
\usepackage{xcolor}
\usepackage{subfigure}
\usepackage[caption=false,font=normalsize,labelfont=sf,textfont=sf]{subfig}
\usepackage{booktabs}
\usepackage{empheq}
\usepackage{balance}

\newcommand{\hdel}[1]{}

\begin{document}
\title{CDiT: Conditional Diffusion Transformer for Geometry-Aware Terahertz Cross Far- and Near-Field Channel Generation\\
}
\author{\IEEEauthorblockN{Zhengdong Hu, and Chong Han,~\IEEEmembership{Senior~Member,~IEEE}}\thanks{
Z. Hu is with the Terahertz Wireless Communications (TWC) Laboratory, Shanghai Jiao Tong University, Shanghai 200240, China (e-mail: {huzhengdong}@sjtu.edu.cn). 

Chong Han is with the Terahertz Wireless Communications (TWC) Laboratory and also the Cooperative Medianet Innovation Center (CMIC), School of Information Science and Electronic Engineering, Shanghai Jiao Tong University, Shanghai 200240, China (e-mail: chong.han@sjtu.edu.cn).
}}
\maketitle
\thispagestyle{empty}
\pdfoptionpdfminorversion=7
\begin{abstract}
Accurate channel modeling is fundamental to design and evaluation of Terahertz (THz) ultra-massive multiple-input multiple-output (UM-MIMO) systems. However, existing model-based approaches typically rely on simplified assumptions, such as sparsity or predefined parametric structures, which are insufficient to capture the complex spatial variations and cross far-/near-field propagation characteristics of practical THz channels. In this paper, a conditional diffusion transformer (CDiT) framework is proposed for high-fidelity THz channel generation. By leveraging the state-of-the-art hybrid planar-spherical wave model (HPSM), THz channel modeling is formulated as a geometry-aware conditional generative learning problem in the sparse beamspace domain. Position information is incorporated as a conditioning signal within a diffusion-transformer architecture, enabling effective learning of the spatially dependent channel distribution. By combining the strong distribution modeling capability of diffusion models with the global dependency modeling strength of transformers, the proposed framework achieves controllable and high-fidelity THz channel synthesis. Extensive experiments on realistic THz channel datasets demonstrate that the proposed framework converges stably and significantly outperforms representative benchmark methods, achieving over 70\% improvement in structural similarity compared with conventional generative adversarial network (GAN)-based approaches, while also consistently surpassing convolutional neural network (CNN)-based diffusion models. Furthermore, the generated channels closely align with the ground-truth angular power distributions, confirming that the proposed framework effectively preserves both the statistical and physically meaningful spatial characteristics of practical THz propagation environments. The proposed framework provides a promising data-driven paradigm for THz channel modeling in next-generation wireless systems.
\end{abstract}

\section{Introduction}

Terahertz (THz) communications are widely recognized as a key enabling technology for next-generation wireless systems due to their abundant spectrum resources and capability to support ultra-high data rates~\cite{thz_survey,thz_survey_josep}. Beyond conventional high-speed wireless transmission, THz communications are expected to play a critical role in emerging applications such as generative AI, where massive real-time data exchange is required to support wireless data centers~\cite{thz_data_center}, immersive online gaming, remote collaboration in embodied intelligence, industrial Internet of Things (IoT), and interactive education services~\cite{zugno2025use}. Moreover, the extremely short wavelength of THz signals enables sub-millimeter spatial resolution, making THz systems highly attractive for localization, target detection, environment reconstruction, and integrated sensing and communication (ISAC)~\cite{6g_thz_isac}. These capabilities are also expected to support future embodied artificial intelligence (AI) systems that require seamless interaction with complex physical environments~\cite{gao2025terahertz, THz_air_ground}. THz communications additionally exhibit strong potential in space-air-ground integrated networks (SAGIN), including high-speed inter-satellite communications, unmanned autonomous platforms, and deep-space information exchange.

Despite these promising advantages, THz communications also encounter significant propagation challenges. The extremely high carrier frequencies result in noticeable free-space path loss and molecular absorption, which considerably attenuate signal power and restrict transmission distances~\cite{channel_survey}. To address these limitations, ultra-massive multiple-input multiple-output (UM-MIMO) systems are typically employed to provide substantial beamforming gains and enhanced spatial multiplexing capability through highly directional transmissions. Consequently, the design, optimization, and performance evaluation of THz UM-MIMO systems critically rely on accurate channel models that can faithfully characterize the underlying propagation environment.

However, channel modeling for THz systems faces several challenges. First, the extremely large antenna arrays lead to high-dimensional channel representations, making accurate modeling computationally demanding. Second, the coexistence of near-field and far-field propagation effects introduces complex spatial characteristics that cannot be accurately captured by conventional planar-wave assumptions~\cite{hpsm_yuhan}. Third, channel statistics exhibit strong dependence on user locations and environmental geometry, which is difficult to characterize using traditional stochastic or geometry-based approaches.

\subsection{Related Work}

Traditional channel modeling is broadly categorized into stochastic, deterministic and hybrid statistical-deterministic approaches. Stochastic approaches often suffer from limited accuracy due to their reliance on idealized probability distributions and empirical parameters \cite{non_stationary_gscm, gscm_2022_thz, indoor_gscm_2020, near_gscm_2026}. For instance, a geometry-based stochastic channel model (GSCM) typically assumes that the scatterer positions follow fixed geometries, such as uniform distributions within circular clusters~\cite{gscm_2022_thz}. In practice, the spatial distribution of scatterers in real-world environments is far more complex, leading to modeling inaccuracies, and such models often fail to account for essential near-field effects. Moreover, such models often fail to account for essential near-field effects. While the refined GSCM proposed in \cite{near_gscm_2026} successfully incorporates near-field propagation via an exact-geometry per-element delay matrix, it prioritizes the 3rd Generation Partnership Project (3GPP) channel model compatibility by intentionally neglecting non-specular effects. Consequently, this framework may under-represent the diffuse multipath components that are physically prevalent in complex THz environments.

In contrast, deterministic methods achieve high modeling accuracy by simulating multipath components, based on fundamental electromagnetic propagation theories \cite{han_rt_2017, rt_thz_yzq_2025, zhang_rt_2015, rt_mag_2024, rt_mag_zjh_2024}. For example, an end-to-end 3D model was proposed to characterize THz channels using ray-tracing (RT), specifically incorporating the effects of graphene-based antenna responses. Furthermore, a site-specific RT model developed in \cite{rt_thz_yzq_2025} was validated against measurements at both 100~GHz and 300~GHz. Despite their precision, deterministic methods suffer from significant computational complexity due to the intensive electromagnetic calculations required, particularly when scaling to electrically large arrays. Moreover, these methods demand exhaustive environmental semantics, including precise geometry and material permittivity profiles, which are often difficult to obtain in practical deployment scenarios.

To balance modeling accuracy and computational efficiency, hybrid statistical-deterministic methods have also been investigated. These approaches combine the site-specific physical interpretability of deterministic RT with the flexibility of statistical modeling. For example, the RT-statistical hybrid model in~\cite{rt_stat_hybrid_twc} integrates RT analysis with statistical characterization of multipath components, achieving better agreement with measured low-THz indoor channels than conventional statistical and GSCM-based models. Nevertheless, hybrid approaches still rely on ray-tracing procedures and environment-dependent calibration, which may limit their scalability when generating large-scale channel datasets across diverse spatial configurations. 

Recently, DL has gained significant prominence and has been widely applied to wireless communications \cite{csi,appro,gan-survey,channel_gan,distribution,cm_dl_2022, cm_dl_survey_2022,cm_digital_2025,cm_nerf_2026,cm_sunsu_2026,cm_wcl_2025,hu_ttgan}. Among various DL architectures, generative adversarial networks (GANs) offer the distinct advantage of accurately modeling complex distributions without requiring prior statistical assumptions. Consequently, GANs have been extensively utilized for channel modeling. For instance, the authors in \cite{appro} trained a GAN to approximate the probability distribution functions (PDFs) of stochastic channel responses, while a GAN-based modeling approach was proposed and validated over AWGN channels in \cite{gan-survey}. Nevertheless, these early works primarily addressed simplified scenarios, necessitating broader applicability to more intricate, practical environments. To tackle increased complexity, the authors in \cite{channel_gan} designed a GAN to generate fake channel samples that accurately mirror the distribution of real samples derived from Clustered Delay Line (CDL) models. Similarly, a model-driven GAN-based approach was developed for Intelligent Reflecting Surface (IRS)-aided systems in \cite{distribution}. These methods typically employ convolutional layers to extract image-like features from channel matrices. 

However, GAN-based channel generation methods often suffer from inherent training instability due to the adversarial optimization process, which may lead to mode collapse and degraded generation diversity. These limitations have motivated the development of more stable generative frameworks, among which diffusion models have recently demonstrated superior generation quality and training robustness across multiple domains \cite{diffusion_beat_gan,dm_survey,song2021scorebased,song2019generative,song2021maximum}. Their potential for wireless channel modeling has also been recently explored in \cite{cm_df_2026,cm_df_tcom_2026}. Specifically, the work in~\cite{cm_df_2026} proposed a denoising diffusion implicit model to learn the relationship between geographic locations and channel characteristics, demonstrating promising generation fidelity and spatial generalization capability. 

Till date, several important challenges remain insufficiently addressed for practical THz UM-MIMO channel modeling. First, existing diffusion-based channel generation approaches mainly focus on far-field propagation scenarios, while the hybrid near-/far-field characteristics introduced by extremely large THz antenna arrays remain largely unexplored. Consequently, conventional planar-wave-based modeling frameworks are unable to accurately characterize the complex cross-field propagation behaviors encountered in practical THz systems. Second, the unprecedentedly high dimensionality of THz UM-MIMO channels and the strong spatial dependence between user geometry and channel structures impose significant challenges on generative modeling. Existing CNN-based diffusion architectures primarily rely on local receptive fields, which limits their ability to effectively capture long-range spatial dependencies and complex global channel characteristics. These limitations motivate the development of a geometry-aware conditional diffusion framework capable of unified hybrid-field channel modeling while effectively learning the spatially-dependent distribution of THz UM-MIMO channels.

\begin{figure*}[t]
    \centering
    \includegraphics[width=0.55\textwidth]{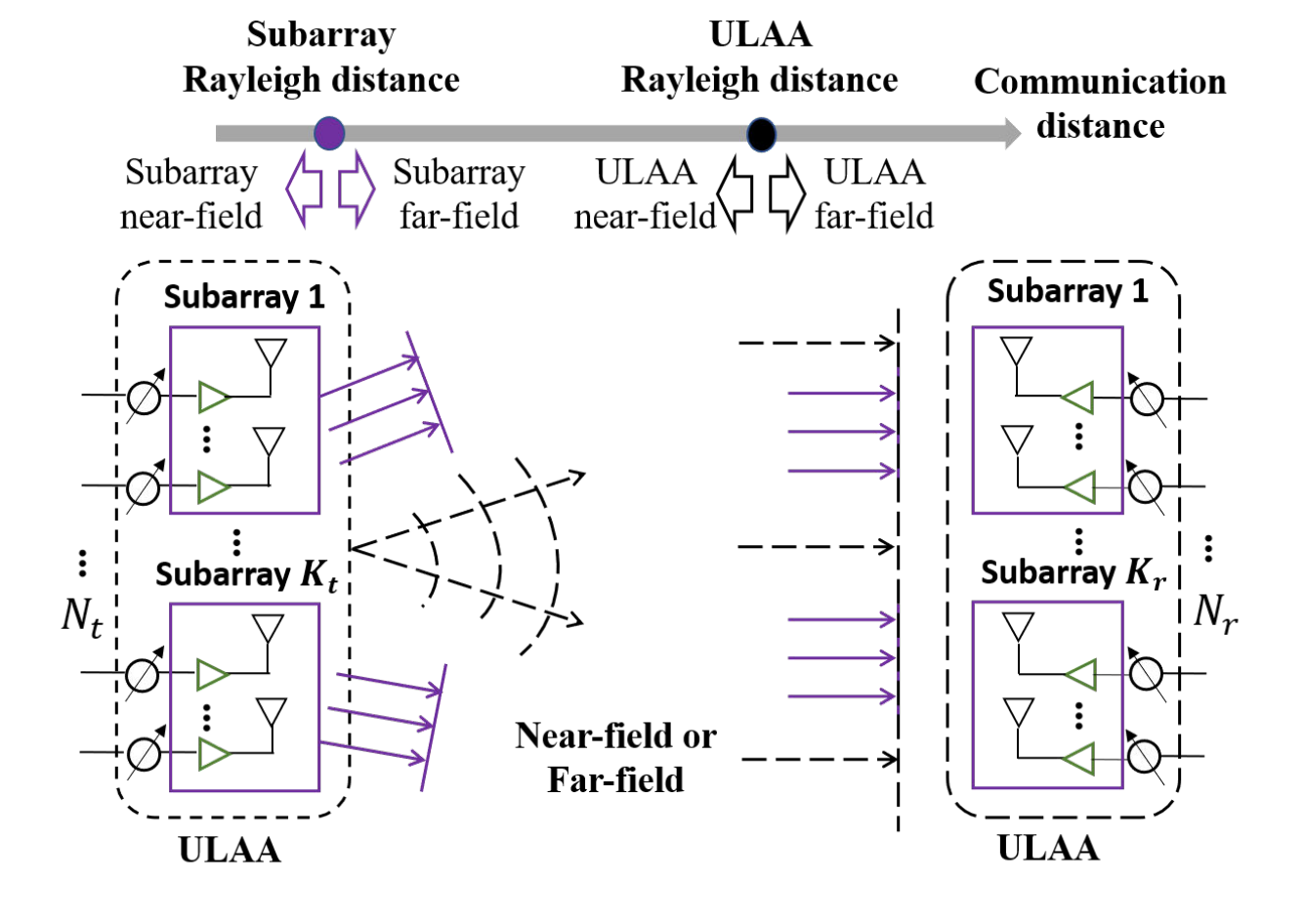}
    \caption{Illustration of THz UM-MIMO system with widely-spaced multi-subarray architecture.}
    \label{fig:sys}
\end{figure*}
\subsection{Contributions}

To address the above challenges, we propose a conditional diffusion transformer (CDiT) framework for geometry-aware THz channel generation. Specifically, the position information is embedded as a conditioning signal within a diffusion transformer (DiT) architecture, enabling the model to jointly capture the spatial dependence and stochastic characteristics of THz channels. Furthermore, we develop a diffusion-based generative framework for unified near-field and far-field channel modeling under the hybrid planar-spherical wave model (HPSM)~\cite{hpsm_yuhan}. By combining the physical interpretability of the HPSM with the strong distribution learning capability of diffusion models, the proposed framework can effectively characterize the complex hybrid-field propagation behaviors of THz UM-MIMO channels.

Compared with conventional CNN-based diffusion models, the proposed transformer-based architecture provides greater flexibility in handling high-dimensional THz channel data and modeling long-range spatial dependencies. In addition, the conditional DiT structure enables efficient fusion between position information and channel features, thereby improving the controllability and spatial consistency of generated channels.

The main contributions of this paper are summarized as follows:

\begin{itemize}

\item \textbf{Unified Hybrid-Field Generative Channel Modeling:}
We develop a diffusion-based generative framework for THz UM-MIMO channel modeling under the HPSM, enabling unified characterization of near-field, far-field, and cross-field propagation behaviors within a single generative model.

\item \textbf{Geometry-Aware Transformer-Based Channel Generation:}
We develop a DiT-based conditional channel generator that incorporates position information as a conditioning signal and leverages global self-attention to capture long-range spatial dependencies across high-dimensional THz UM-MIMO channels. This design enables controllable generation of THz channel realizations under different spatial configurations while improving the preservation of dominant propagation clusters, angular power distributions, and spatial channel coherence.

\item \textbf{Experimental Validation of Channel Fidelity:}
Experimental results show that the generated channels exhibit strong agreement with ground-truth channels in terms of both structural similarity and angular power distribution. The proposed framework also outperforms representative baseline methods, demonstrating its effectiveness in preserving dominant propagation structures and physically meaningful angular energy distributions.

\end{itemize}

The remainder of this paper is organized as follows. In Section~\ref{sec:sys}, the THz UM-MIMO system model is introduced, where the HPSM channel model is described in detail. Section~\ref{sec:method} presents the proposed CDiT based channel modeling framework. The performance of the proposed method is evaluated in Section~\ref{sec:evaluation}. Finally, Section~\ref{sec:summary} concludes this paper.

\section{System Model and Problem Formulation}\label{sec:sys}
In this section, we first describe the channel model and its cross far- and near-field characteristics using HPSM channel model. Then, we formulate the channel modeling task as a conditional channel generation problem, where channel realizations are generated conditioned on spatial position information.

\subsection{System Model}
We consider a THz UM-MIMO system as depicted in Fig.~\ref{fig:sys}, where large-scale antenna arrays are deployed at both the transmitter (Tx) and receiver (Rx) to compensate for severe propagation loss at high frequencies. Specifically, the Tx and Rx are equipped with $N_t$ and $N_r$ antennas, respectively, organized into $K_t$ and $K_r$ subarrays following a widely-spaced multi-subarray architecture. Each subarray contains closely spaced antenna elements, while the spacing between subarrays is significantly larger, enabling enhanced spatial multiplexing and improved beamforming capability.

Due to the extremely high carrier frequencies and ultra-large antenna array apertures, THz wireless channels exhibit distinctive propagation characteristics. The transition between near-field and far-field propagation regions is determined by the Rayleigh distance, which increases quadratically with the array aperture and decreases with the carrier wavelength~\cite{rayleigh_add}. As illustrated in Fig.~\ref{fig:sys}, the antenna array of Rx may reside in either the near-field or far-field region of the Tx antenna array depending on the transmission distance. Consequently, practical THz propagation environments often involve the coexistence of near-field and far-field effects, resulting in hybrid wavefront propagation behaviors that cannot be accurately characterized by conventional planar-wave-based channel models.

In the far-field regime, the channel is commonly modeled using the planar wave model (PWM)~\cite{pwm_swm}, where each propagation path is characterized by a complex gain and angular parameters. The channel matrix can be expressed as
\begin{equation}
\mathbf{H}_{\text{PWM}} = \sum_{l=1}^{L} \alpha_l \mathbf{a}_r(\theta_{r,l}, \phi_{r,l}) \mathbf{a}_t^H(\theta_{t,l}, \phi_{t,l}),
\end{equation}
where $L$ represents the number of propagation paths, $\alpha_l$ denotes the complex path gain, and $\mathbf{a}_r(\theta_{r,l}, \phi_{r,l}), \mathbf{a}_t(\theta_{t,l}, \phi_{t,l})$ represent the array response vectors at the Rx and Tx, respectively.

In the near-field regime, the spherical wave model (SWM) must be adopted to account for distance-dependent phase variations across antenna elements~\cite{pwm_swm}. Unlike the far-field approximation, the channel response depends explicitly on the unique propagation distance between every transmit-receive antenna pair, resulting in significantly higher modeling complexity. Specifically, the channel response between the $i^{\mathrm{th}}$ transmit antenna and $n^{\mathrm{th}}$ receive antenna is expressed as
\begin{equation}
    \mathbf{H}_{\text{SWM}}[i,n] = \sum_{l=1}^{L} |\alpha^{i,n}_l| e^{-j \frac{2\pi}{\lambda} d^{i,n}_l},
\end{equation}
where $l = 1, \dots, L$ denotes the index of propagation paths between the $i^{\mathrm{th}}$ Tx and $n^{\mathrm{th}}$ Rx antennas, $|\alpha^{i,n}_l|$ represents the magnitude of the path gain for the $l^{\mathrm{th}}$ path, and $d^{i,n}_l$ is the corresponding propagation distance. The phase term $e^{-j \frac{2\pi d^{i,n}_l}{\lambda}}$ captures the phase shift associated with the distance-dependent propagation.

For each antenna pair $(i, n)$, the SWM necessitates $2L$ parameters $\{|\alpha^{i,n}_l|, d^{i,n}_l\}$ to fully describe the channel response. Consequently, the total number of parameters in the complete Tx-Rx channel matrix scales as $2LN_rN_t$, which becomes prohibitively large for THz UM-MIMO systems due to their extreme array dimensions. By contrast, the PWM offers a much more efficient representation, but its planar-wave assumption neglects the distance-dependent phase evolution across antenna elements and therefore becomes inaccurate in near-field and hybrid-field propagation scenarios. Consequently, neither model alone provides a satisfactory solution for practical THz UM-MIMO channel modeling, thereby motivating the adoption of hybrid representations that balance physical fidelity and computational tractability.

\subsection{HPSM Channel Model}
To balance modeling accuracy and computational complexity, we adopt the HPSM channel model, which provides a unified framework for characterizing both near-field and far-field propagation effects. The key idea of HPSM is to exploit the structural properties of the multi-subarray architecture. Specifically, within each subarray, the antenna aperture is relatively small, and the propagation can be well approximated by planar wavefronts. However, across different subarrays, the large inter-subarray spacing leads to noticeable spherical wave effects.

Based on this observation, the channel between the $k_t$-th transmit subarray and the $k_r$-th receive subarray can be expressed as
\begin{equation}\label{hpsm}
\begin{aligned}
    \mathbf{H}_{\text{HPSM}}[k_t,k_r]=&\sum_{l=1}^L|\alpha_l^{k_t,k_r}|e^{-j\phi_l^{k_t,k_r}}\\
    &\mathbf{a}_{k_r}(\theta_{k_r,l}, \phi_{k_r,l})
    \mathbf{a}_{k_t}^H(\theta_{k_t,l}, \phi_{k_t,l}),
\end{aligned}
\end{equation}
where $|\alpha_l^{k_t,k_r}|$ and $\phi_l^{k_t,k_r}$ denote the magnitude and phase shift of the $l$th propagation path between the $k_t$th Tx subarray and the $k_r$th Rx subarray, respectively. The vectors $\mathbf{a}_{k_r}(\theta_{k_r,l}, \phi_{k_r,l})$ and $\mathbf{a}_{k_t}(\theta_{k_t,l}, \phi_{k_t,l})$ denote the corresponding Rx and Tx subarray response vectors, parameterized by the angles of arrival (AoA) $(\theta_{k_r,l}, \phi_{k_r,l})$ and angles of departure (AoD) $(\theta_{k_t,l}, \phi_{k_t,l})$, respectively.

The HPSM not only balances modeling accuracy and computational complexity, but also admits a structured beamspace representation that is suitable for generative learning. In the conventional far-field regime, the channel matrix $\mathbf{H}_P$ can be represented in the angular domain as
\begin{equation}
\mathbf{H}_{\text{PWM}} = \mathbf{A}_R \mathbf{H}_b \mathbf{A}_T^H,
\end{equation}
where $\mathbf{A}_R \in \mathbb{C}^{N_r \times N_r}$ and $\mathbf{A}_T \in \mathbb{C}^{N_t \times N_t}$ denote the 2D discrete Fourier transform (DFT)-based array response dictionaries at the Rx and Tx, respectively, and $\mathbf{H}_b$ is the corresponding beamspace channel matrix. Due to the limited number of dominant propagation paths in THz channels, $\mathbf{H}_b$ typically exhibits a sparse or approximately sparse structure.

A similar beamspace representation can be constructed for the HPSM framework~\cite{hpsm_yuhan}. Since the intra-subarray channel responses are modeled using PWM, the corresponding subarray channels admit sparse angular-domain representations. Specifically, suppose that the Rx array is partitioned into $K_r$ subarrays and the Tx array into $K_t$ subarrays, where each Rx subarray contains $N_r^{\mathrm{sub}} = N_r / K_r$ antennas and each Tx subarray contains $N_t^{\mathrm{sub}} = N_t / K_t$ antennas, assuming that $N_r$ and $N_t$ are divisible by $K_r$ and $K_t$, respectively.

The corresponding subarray beamspace dictionaries are denoted by $\mathbf{A}_{k_r} \in \mathbb{C}^{N_r^{\mathrm{sub}} \times N_r^{\mathrm{sub}}}$ for the Rx and $\mathbf{A}_{k_t} \in \mathbb{C}^{N_t^{\mathrm{sub}} \times N_t^{\mathrm{sub}}}$ for the Tx. Accordingly, the overall Rx beamspace dictionary is constructed as the block-diagonal matrix

\begin{equation}
\overline{\mathbf{A}}_R
=
\mathrm{blkdiag}
\left[
\mathbf{A}_{1}, \ldots, \mathbf{A}_{K_r}
\right].
\end{equation}

Similarly, the Tx beamspace dictionary $\overline{\mathbf{A}}_T \in \mathbb{C}^{N_t \times N_t}$ is constructed in the same manner. Based on these structured dictionaries, the HPSM channel model in \eqref{hpsm} can be approximately represented in the beamspace domain as

\begin{equation}\label{beamrep}
\mathbf{H}_{\mathrm{HPSM}}
\approx
\overline{\mathbf{A}}_R
\overline{\mathbf{H}}_b
\overline{\mathbf{A}}_T^{\mathsf{H}},
\end{equation}
where $\overline{\mathbf{H}}_b \in \mathbb{C}^{N_r \times N_t}$ denotes the beamspace channel representation. Although practical THz channels may not exhibit strict sparsity due to scattering, leakage, and hybrid-field coupling effects, this transformation provides a highly structured and approximately sparse representation, which substantially reduces the complexity of the generative modeling task and facilitates more effective distribution learning.

\subsection{Problem Formulation}
The objective of THz channel modeling is to accurately characterize the conditional distribution of channel realizations under varying spatial configurations. Let $\mathbf{p} \in \mathbb{R}^{8}$ denote the geometry-aware conditioning vector, defined as

\begin{equation}
\mathbf{p}
=
[d, x, y, z, \sin(\theta), \cos(\theta), \sin(\phi), \cos(\phi)]^T,
\end{equation}
where $d$ denotes the Euclidean distance between the Tx and Rx, $(x,y,z)$ represents the relative Cartesian coordinates of the Rx with respect to the Tx, and $\theta$ and $\phi$ denote the azimuth and elevation angles from the Tx to the Rx, respectively. To avoid discontinuities caused by angular periodicity, the directional information is encoded using sine and cosine transformations, thereby providing a continuous geometric representation for the proposed conditional generative framework.

Rather than directly modeling the high-dimensional complex channel matrix $\mathbf{H}$, we leverage the HPSM to learn the conditional distribution of its sparse beamspace representation, denoted by $\overline{\mathbf{H}}_b$. 
This beamspace representation is adopted for three main reasons. First, THz channels typically exhibit strong sparsity in the angular domain due to the limited number of dominant propagation paths, resulting in a more structured and compact representation than the raw spatial-domain channel matrix. Second, learning the distribution of a sparse representation significantly reduces the complexity of the generative modeling task, allowing the diffusion model to focus on the dominant propagation structure rather than highly fluctuating dense channel coefficients. Third, since the beamspace representation is derived from the physically interpretable HPSM framework, learning in this domain preserves the underlying propagation structure while improving modeling efficiency.

Based on this representation, THz channel modeling is formulated as a conditional generative learning problem. Specifically, the objective is to learn a parameterized conditional generative model $G_{\theta}$ that transforms a latent Gaussian variable into a beamspace channel realization conditioned on the geometry information:

\begin{equation}
\overline{\mathbf{H}}_b
=
G_{\theta}(\mathbf{p},\mathbf{z}),
\quad
\mathbf{z}
\sim
\mathcal{N}(\mathbf{0},\mathbf{I}),
\end{equation}
such that the generated samples follow the conditional distribution

\begin{equation}
\overline{\mathbf{H}}_b
\sim
p(\overline{\mathbf{H}}_b|\mathbf{p}).
\end{equation}

This formulation transforms THz channel modeling from a conventional model-driven characterization problem into a geometry-aware generative learning framework. As a result, the proposed approach enables the synthesis of realistic and controllable channel realizations for arbitrary spatial configurations while preserving the stochastic characteristics of practical wireless propagation environments.
\section{Conditional Diffusion Transformer for THz Channel Generation}\label{sec:method}
In this section, we present the proposed CDiT framework for geometry-aware THz channel generation. The proposed framework synergistically combines the robust distribution learning capabilities of diffusion models with the global dependency modeling power of transformers, enabling effective learning of the conditional channel distribution, specifically the sparse beamspace representation $\overline{\mathbf{H}}_b$, under complex hybrid near-/far-field propagation environments. 

To achieve this, the CDiT framework is structured into two core components. First, a conditional diffusion process provides the stochastic foundation for perturbing THz channel samples into Gaussian noise and subsequently synthesizing clean channel data through a reverse denoising process conditioned on spatial geometry. Second, a specialized DiT-based denoising network serves as the backbone of the architecture, leveraging multi-head self-attention mechanisms to capture long-range spatial dependencies and reconstruct high-fidelity channel structures from noisy observations

\begin{figure*}[t]
    \centering
    \includegraphics[width=1.0\textwidth]{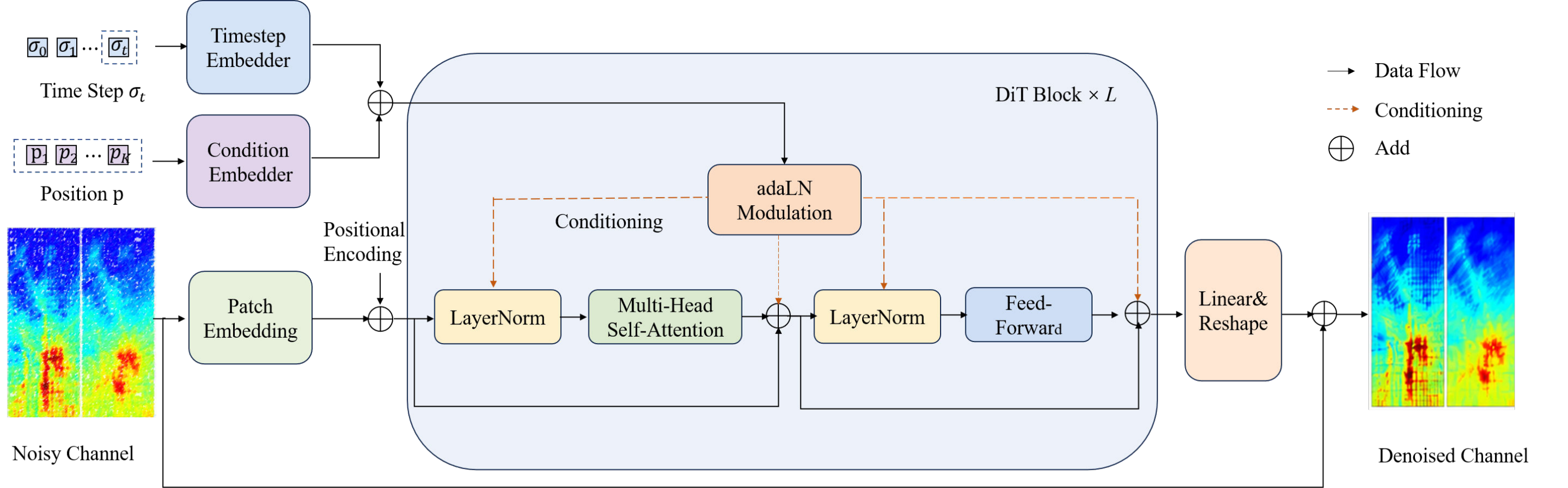}
    \caption{Framework of the conditional diffusion transformer structure.}
    \label{fig_model}
\end{figure*}

\subsection{Conditional Diffusion Process}
To model the conditional distribution of THz channels under varying spatial configurations, we adopt a diffusion-based generative framework that progressively transforms a simple Gaussian distribution into the target THz channel distribution. Compared with GAN, diffusion models generally provide more stable optimization and stronger distribution learning capability, making them particularly suitable for high-dimensional THz channel generation.

The clean THz channel representation is denoted by $\mathbf{H}_0 \in \mathbb{R}^{2\times N_r\times N_t}$, where the two channels correspond to the real and imaginary components of the complex channel matrix. Given the conditional channel distribution $p(\mathbf{H}_0|\mathbf{p})$, where $\mathbf{c}$ denotes the geometry-aware conditioning information, the forward diffusion process progressively perturbs the channel sample by injecting Gaussian noise with monotonically increasing noise levels, i.e., $\sigma_{t_1}^2 < \sigma_{t_2}^2 < \cdots < \sigma_{t_K}^2$, over the continuous time interval $t \in [0,T]$. Through this process, the structured THz channel distribution is gradually transformed into an isotropic Gaussian distribution.

Following the diffusion formulation, the forward perturbation process is modeled by the stochastic differential equation (SDE)

\begin{equation}
\mathrm{d}\mathbf{H}_t
=
\sqrt{\frac{\mathrm{d}\sigma_t^2}{\mathrm{d}t}}
\,\mathrm{d}\mathbf{w},
\end{equation}
where $\mathbf{w}$ denotes a standard Wiener process and $\sigma_t$ defines the noise schedule controlling the perturbation intensity. In this work, a linear noise schedule with $\sigma(t)=t$ is adopted, which has demonstrated effective performance for modeling the sparse and highly structured distribution of THz channels.

As the forward diffusion process evolves, the original channel sample is progressively corrupted into Gaussian noise, whereas the reverse process seeks to recover the clean channel distribution from noisy observations. Rather than directly solving the reverse-time stochastic process, we adopt the probability flow ordinary differential equation (ODE), which provides a deterministic sampling trajectory with improved inference efficiency. The reverse diffusion process is formulated as

\begin{equation}\label{ode}
\mathrm{d}\mathbf{H}_t
=
-
\dot{\sigma}_t \sigma_t
\nabla_{\mathbf{H}_t}
\log p(\mathbf{H}_t|\mathbf{p})
\,\mathrm{d}t,
\end{equation}
where $\dot{\sigma}_t$ denotes the derivative of the noise schedule, and $\nabla_{\mathbf{H}_t}\log p(\mathbf{H}_t|\mathbf{p})$ represents the conditional score function, which points toward regions of higher probability density under the geometry-conditioned channel distribution.

During inference, the reverse ODE is numerically solved using the Euler method
\begin{equation}\label{solver}
\mathbf{H}_{t-\Delta t}
=
\mathbf{H}_t
+
\Delta t
\,\dot{\sigma}_t \sigma_t
\nabla_{\mathbf{H}_t}
\log p(\mathbf{H}_t|\mathbf{p}),
\end{equation}
where $\Delta t$ denotes the discretization step size. To realize this reverse sampling process, the conditional score function must be estimated by a neural network. Instead of directly parameterizing the score function, we model the network as a denoiser $\mathbf{D}_{\theta}(\mathbf{H}_t,\sigma_t,\mathbf{p})$, which predicts the clean channel sample from the noisy observation under the given geometry condition. The network parameters are optimized by minimizing the denoising objective

\begin{equation}\label{eq_loss}
\theta^*
=
\arg\min_{\theta}
\mathbb{E}_{\mathbf{H}_0,\mathbf{p},\sigma_t}
\left[
\left\|
\mathbf{D}_{\theta}(\mathbf{H}_t,\sigma_t,\mathbf{p})
-
\mathbf{H}_0
\right\|_2^2
\right].
\end{equation}

Once trained, the corresponding score function can be recovered as

\begin{equation}
\mathbf{s}_{\theta}(\mathbf{H}_t,\sigma_t,\mathbf{p})
=
\frac{
\mathbf{D}_{\theta}(\mathbf{H}_t,\sigma_t,\mathbf{p})
-
\mathbf{H}_t
}{
\sigma_t^2
}.
\end{equation}

This learned score function guides the reverse diffusion process, enabling the proposed framework to progressively reconstruct realistic THz channel realizations from Gaussian noise while preserving the geometry-dependent propagation characteristics imposed by the position condition.

\subsection{DiT-Based Conditional Denoising Network}

To effectively model the high-dimensional spatial characteristics and long-range dependencies inherent in THz UM-MIMO channels, we adopt a diffusion transformer (DiT) as the denoising backbone of the proposed framework. Compared with conventional CNN-based diffusion architectures, which primarily rely on localized receptive fields, transformer architectures leverage self-attention mechanisms to directly capture global interactions across the entire channel representation. This capability is particularly valuable for THz channel generation, where the propagation structure is governed not only by local multipath sparsity but also by long-range spatial correlations arising from ultra-large antenna apertures and hybrid near-/far-field propagation effects. By enabling direct information exchange between distant channel regions, the transformer backbone is well suited for modeling the complex global spatial dependencies inherent in THz UM-MIMO channels.

The overall architecture of the proposed conditional DiT framework is illustrated in Fig.~\ref{fig_model}. The network takes as input the noisy channel representation $\mathbf{H}_t \in \mathbb{R}^{2\times N_r \times N_t}$ at diffusion timestep $t$, together with the position information $\mathbf{p}$. The two channels of $\mathbf{H}_t$ correspond to the real and imaginary components of the complex THz channel matrix, respectively. This two-channel tensor representation enables the proposed framework to naturally leverage transformer-based visual modeling while preserving the intrinsic spatial structure of the wireless channel. The conditioning information and diffusion timestep are embedded into a unified conditioning representation, which is injected into the transformer backbone to guide the denoising process.

\subsubsection{Patch Embedding and Positional Encoding}

Following the standard DiT architecture, the noisy channel input is first converted into a sequence of latent tokens through a patch embedding module. Specifically, the input channel tensor is partitioned into non-overlapping patches of size $P \times P$, where each patch is flattened and linearly projected into a $D$-dimensional latent embedding space. The resulting token sequence is represented as

\begin{equation}
\mathbf{X}
=
[\mathbf{x}_1,\mathbf{x}_2,\ldots,\mathbf{x}_N]
\in
\mathbb{R}^{N\times D},
\end{equation}
where the total number of tokens is given by $N = N_rN_t/P^2$.

This patch-based tokenization significantly reduces the sequence length compared with pixel-level representations, thereby improving the computational efficiency of the self-attention mechanism while preserving the dominant structural characteristics of the channel matrix.

Since transformers do not inherently encode spatial ordering information, fixed two-dimensional sinusoidal positional embeddings are added to the token sequence to preserve positional awareness. This encoding enables the transformer to distinguish the relative locations of different channel regions, which is particularly important for THz channel modeling, where antenna-dependent propagation characteristics and hybrid near-/far-field spatial structures must be accurately preserved during the generative process.

\subsubsection{Conditional Embedding Mechanism}

To enable geometry-aware channel generation, the proposed framework incorporates both the diffusion timestep and user position information as conditioning signals.

The diffusion timestep $t$, which reflects the current noise level in the reverse diffusion process, is encoded using a sinusoidal timestep embedding followed by a multi-layer perceptron (MLP), yielding the latent timestep representation $\mathbf{e}_t \in \mathbb{R}^{D}$. Meanwhile, the user position vector $\mathbf{p}$, which characterizes the spatial geometry of the communication environment, is processed through a continuous conditioning embedding network to obtain the geometry-aware embedding $\mathbf{e}_p \in \mathbb{R}^{D}$.

The final conditioning representation is formed by combining the two embeddings as

\begin{equation}
\mathbf{c}
=
\mathbf{e}_t + \mathbf{e}_p.
\end{equation}

Rather than adopting token-concatenation-based conditioning strategies, the proposed framework follows the standard conditional DiT design, in which the conditioning representation is injected into the transformer backbone through adaptive layer normalization (adaLN). This mechanism enables dynamic modulation of the internal feature representations according to both the denoising stage and the underlying spatial geometry, thereby facilitating controllable and geometry-aware THz channel generation.

\subsubsection{Transformer Denoising Backbone}
The denoising backbone consists of multiple stacked DiT blocks, each comprising multi-head self-attention (MHSA), adaptive normalization layers, feed-forward networks (FFNs), and residual connections.

Within each DiT block, the input token sequence is first normalized and modulated by the conditioning vector $\mathbf{c}$ through adaptive layer normalization (adaLN), expressed as

\begin{equation}
\mathrm{adaLN}(\mathbf{x},\mathbf{c})
=
\gamma(\mathbf{c}) \odot
\mathrm{LayerNorm}(\mathbf{x})
+
\beta(\mathbf{c}),
\end{equation}
where $\gamma(\cdot)$ and $\beta(\cdot)$ denote learnable scaling and shifting functions generated from the conditioning representation, respectively.

This conditioning mechanism allows the geometric information and diffusion timestep to directly influence feature transformation at every transformer layer, rather than being incorporated only at the input stage. As a result, the denoising process can be dynamically adapted to different spatial configurations and noise levels.

The modulated token sequence is subsequently processed by the MHSA module, which captures long-range dependencies through global pairwise interactions among all channel patches. The attention operation is defined as

\begin{equation}
\mathrm{Attention}(\mathbf{Q},\mathbf{K},\mathbf{V})
=
\mathrm{Softmax}
\left(
\frac{\mathbf{Q}\mathbf{K}^{T}}{\sqrt{D}}
\right)
\mathbf{V},
\end{equation}
where $\mathbf{Q}$, $\mathbf{K}$, and $\mathbf{V}$ denote the query, key, and value matrices derived from the token embeddings.

Compared with CNN-based architectures, where information exchange is restricted by local convolutional receptive fields, self-attention enables direct interaction between all token pairs. This capability is particularly beneficial for THz UM-MIMO channel modeling, where hybrid near-/far-field propagation and ultra-large antenna apertures introduce complex spatial dependencies spanning the entire channel representation.

Following the attention operation, an FNN further enhances nonlinear feature transformation, while residual connections are employed throughout the transformer backbone to improve gradient propagation and stabilize deep network training.

\subsubsection{Output Reconstruction}

After passing through the stacked transformer blocks, the processed latent tokens are projected back to the original channel space through a final linear projection layer. Specifically, each latent token is mapped to its corresponding patch representation, followed by an unpatchify operation that reconstructs the complete channel tensor $\hat{\mathbf{H}}_0 \in \mathbb{R}^{2\times H\times W}$.

In addition, the proposed framework adopts a preconditioned denoising formulation, which introduces an implicit residual connection between the noisy input and the network output. This design allows the model to focus on learning the residual correction required to recover the clean channel realization, rather than reconstructing the entire channel structure from scratch. Such a residual learning strategy improves optimization stability, accelerates convergence, and enhances reconstruction fidelity, particularly for high-dimensional THz channel representations.

The reconstructed output corresponds to the predicted denoised THz channel sample, where the two channels represent the real and imaginary components of the complex channel matrix. Through this end-to-end conditional denoising architecture, the proposed framework progressively reconstructs realistic THz channel realizations from noisy observations while preserving both the statistical characteristics and spatial propagation structures of practical wireless environments.
\section{Results and Performance Evaluation}\label{sec:evaluation}

In this section, we evaluate the performance of the proposed CDiT framework for THz channel generation. First, the experimental setup is introduced, including the dataset generation procedure and the training configurations of the proposed network. Next, the convergence behavior of the proposed framework is investigated through the training and testing loss curves. Finally, the channel modeling performance is evaluated by comparing the generated channels with the ground-truth channels in terms of channel similarity, spatial characteristics, and power spectrum distributions.

\begin{figure}
    \centering
    \includegraphics[width=0.5\textwidth]{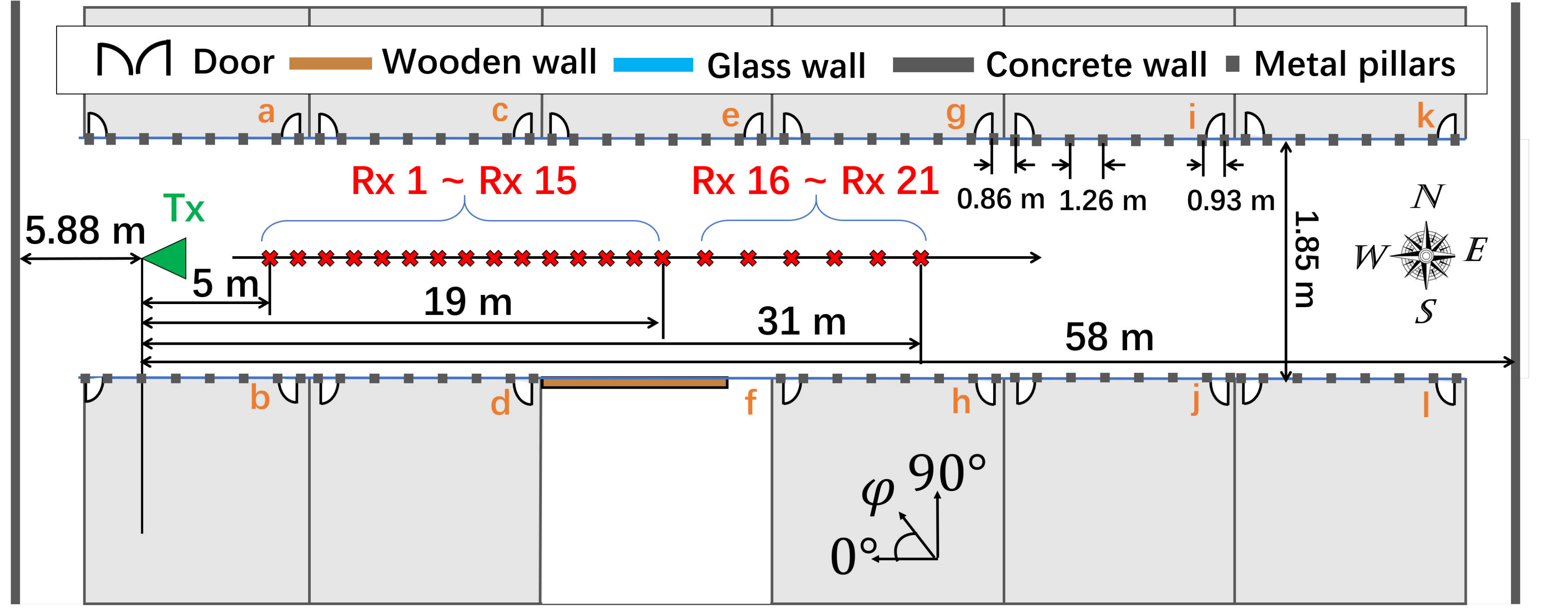}
    \caption{Measurement layout in the indoor corridor scenario.}
    \label{fig_simulation}
\end{figure}
\subsection{Experimental Setup}

In this subsection, we describe the THz channel dataset generation procedure, the training configuration of the proposed CDiT framework, and the hardware platform used for performance evaluation.

\subsubsection{THz Channel Dataset Generation}

The THz channel dataset is generated using QuaDRiGa~\cite{quadriga}, calibrated with real measurement statistics obtained from indoor corridor measurements at 306--321 GHz conducted by our group~\cite{yuanbo_icc}. The considered THz UM-MIMO system consists of $N_t = 256$ transmit antennas and $N_r = 64$ receive antennas, partitioned into $K_t = 2$ and $K_r = 2$ subarrays at the transmitter and receiver, respectively, to support the HPSM-based hybrid near-/far-field channel representation. In total, 50,000 channel realizations are synthesized, with 45,000 samples used for training and the remaining 5,000 reserved for testing.

To ensure realistic propagation characteristics, the calibration process incorporates key statistical parameters extracted from the measurement campaign, including the path loss exponent, the mean and standard deviation of the Rician $K$-factor, delay spread, angular spread, and spatial correlation characteristics. These measured statistics are integrated into the geometry-based stochastic channel model (GSCM) implemented in QuaDRiGa, enabling the synthesized multipath propagation characteristics to closely resemble practical THz wireless environments.

Furthermore, the generated channels follow the hybrid planar-spherical wave model (HPSM), allowing simultaneous characterization of near-field and far-field propagation effects in THz UM-MIMO systems. The considered indoor corridor measurement scenario is illustrated in Fig.~\ref{fig_simulation}.

\subsubsection{Training Configuration}

The proposed CDiT framework is trained using the score-based diffusion objective described in Section III. During training, the complex THz channel matrix is represented as a two-channel real-valued tensor, where the two channels correspond to the real and imaginary components, respectively.

To improve training stability, the channel samples are normalized using Frobenius norm-based scaling prior to training. The input channel representation has a dimension of $2 \times N_r \times N_t$. The denoising backbone adopts the proposed conditional DiT architecture, in which the noisy channel sample is partitioned into non-overlapping patches and projected into latent embeddings. User position information and diffusion timestep information are incorporated as conditioning signals through the adaptive layer normalization mechanism.

Model optimization is performed using the Adam optimizer with an initial learning rate of $10^{-4}$, momentum coefficient $\beta_1=0.9$, and numerical stabilization factor $\epsilon=10^{-3}$. To further improve optimization stability and generation robustness, exponential moving average (EMA) is employed with a decay factor of $0.999$.

The network is trained for 100 epochs using distributed data parallel (DDP) training with a batch size of 8 per GPU. To evaluate spatial generalization capability, the testing dataset contains channel realizations associated with user locations and propagation conditions distinct from those observed during training.

\subsubsection{Hardware Platform}

All experiments are conducted on a high-performance workstation equipped with an INTEL(R) XEON(R) GOLD 6530 Processor and eight NVIDIA GeForce RTX 4090 GPUs. The multi-GPU distributed training framework enables efficient acceleration of the computationally intensive diffusion-based channel generation task.

\subsection{Convergence Evaluation}

\begin{figure}
    \centering
    \includegraphics[width=0.5\textwidth]{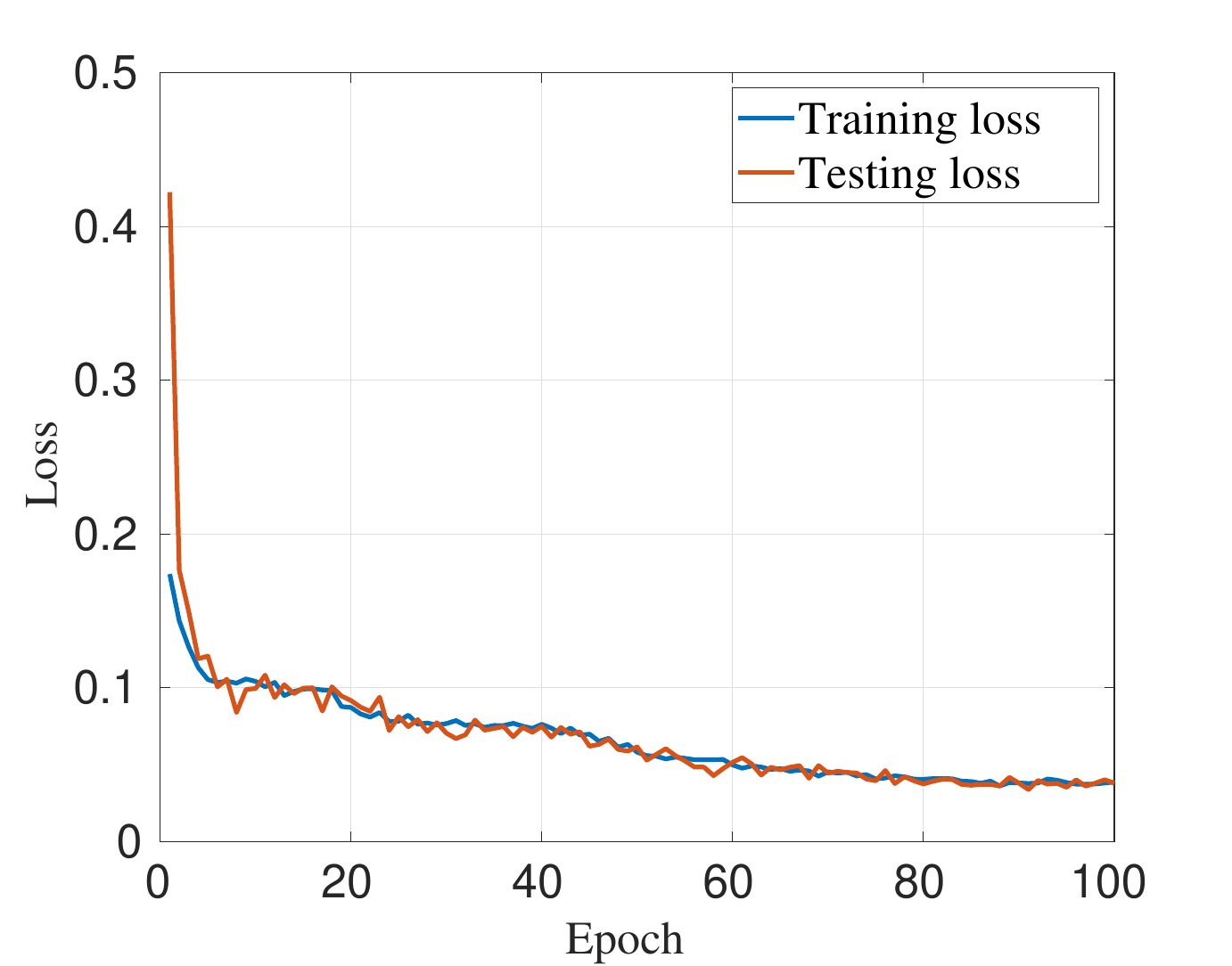}
    \caption{Training loss and testing loss versus number of epochs.}
    \label{fig_loss}
\end{figure}

The training and testing loss curves of the proposed CDiT framework are presented in Fig.~\ref{fig_loss}, where each epoch corresponds to one complete pass through the training dataset.

Both loss curves exhibit a rapid decrease during the early stage of training, indicating that the proposed diffusion framework efficiently captures the underlying THz channel distribution. The testing loss gradually stabilizes after approximately 60 epochs, suggesting satisfactory convergence of the generative model.

Minor fluctuations can be observed in the training loss throughout the optimization process, primarily due to the stochastic nature of mini-batch gradient optimization and the random noise perturbations inherent to diffusion-based learning. In contrast, the testing loss exhibits a consistently decreasing and stable trend, indicating that the exponential moving average (EMA) strategy effectively improves optimization stability and mitigates parameter oscillations.

These observations confirm that the proposed CDiT framework achieves stable convergence and demonstrates strong learning capability for high-dimensional THz channel generation.

 \begin{figure*}[t]
		\centering
		\subfigure[Original channel.
		]{\includegraphics[width=0.4\textwidth]{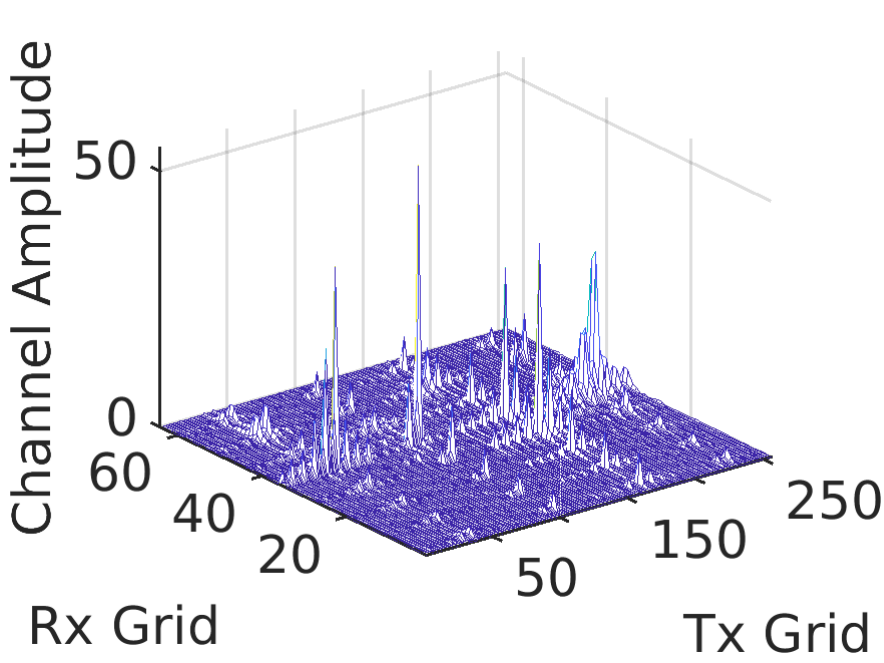}}
		\subfigure[Generated channel.]{\includegraphics[width=0.4\textwidth]{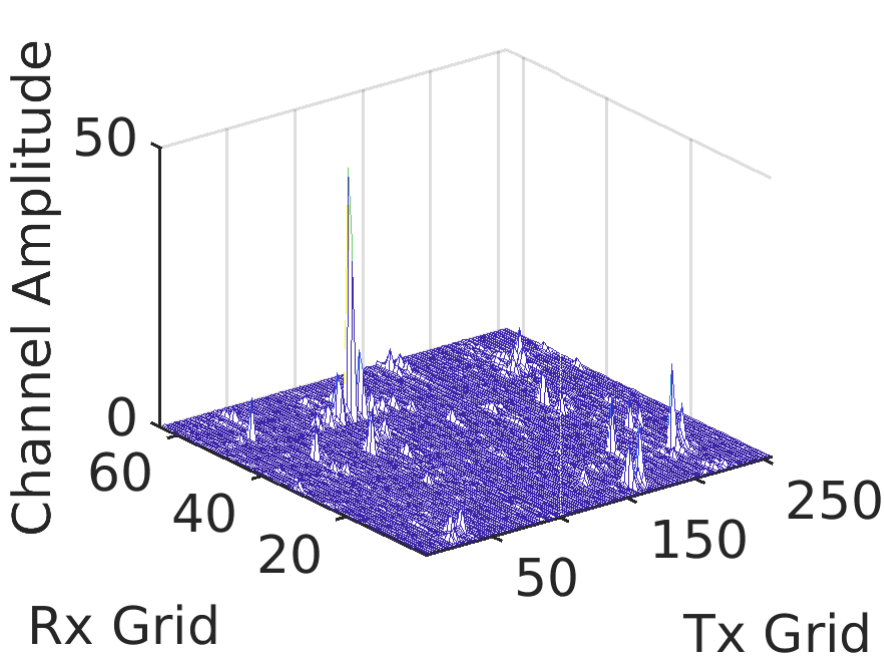}} 
		\caption{Plot of original and generated channel.}
		\label{fig_structure} 
\end{figure*}

\begin{figure}
    \centering
    \includegraphics[width=0.5\textwidth]{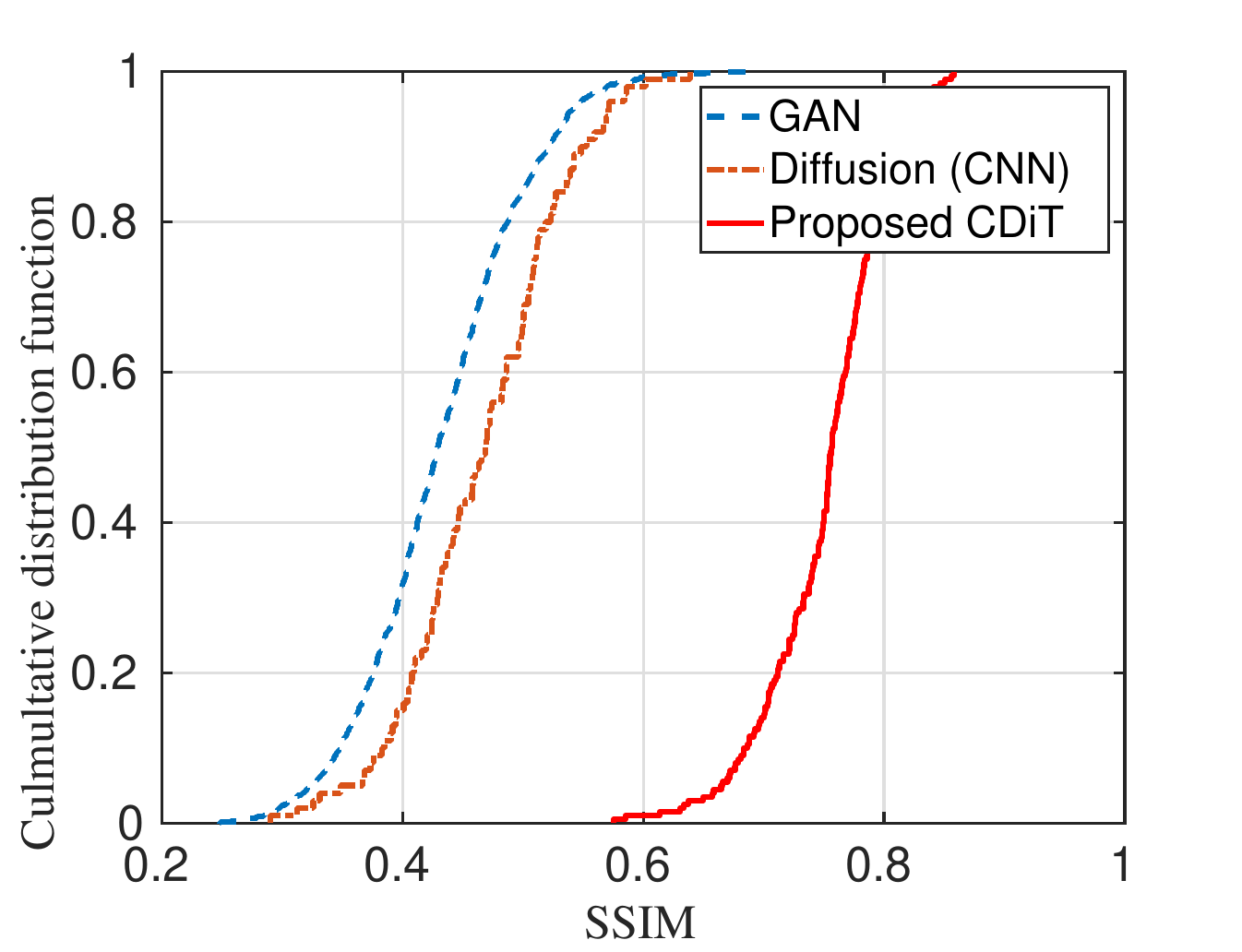}
    \caption{SSIM of channel matrix for the generated channels.}
    \label{fig_ssim}
\end{figure}

\subsection{Structural Similarity Evaluation}
Representative ground-truth and generated THz channel realizations at 0.3 THz are presented in Fig.~\ref{fig_structure}. It can be observed that the channels generated by the proposed CDiT framework closely resemble the corresponding ground-truth samples, preserving both the dominant multipath components (MPCs) and the overall angular-domain structural patterns. This visual consistency suggests that the proposed generative framework effectively captures the intrinsic spatial structure and statistical characteristics of practical THz propagation environments.

To quantitatively evaluate the structural similarity between generated and ground-truth channels, the structural similarity index measure (SSIM) is employed. SSIM is a widely used perceptual similarity metric that jointly evaluates luminance, contrast, and structural consistency between two samples~\cite{ref_ssim}. Its value ranges from 0 to 1, with larger values indicating higher structural similarity. For each channel realization, the generated channel matrix is compared with its corresponding ground-truth sample, and the cumulative distribution function (CDF) of the resulting SSIM values is shown in Fig.~\ref{fig_ssim}.

To further assess the effectiveness of the proposed framework, two representative benchmark methods are considered for comparison: a GAN-based channel generation model and a CNN-based diffusion model. The GAN benchmark represents conventional adversarial generative modeling, while the CNN-based diffusion benchmark isolates the impact of replacing the transformer denoising backbone with a convolutional architecture.

As shown in Fig.~\ref{fig_ssim}, the proposed CDiT framework consistently outperforms both benchmark methods across the entire SSIM distribution. Specifically, the average SSIM values achieved by the GAN-based model, CNN-based diffusion model, and the proposed CDiT framework are 0.43, 0.46, and 0.75, respectively. This corresponds to an improvement of approximately 74.4\% over the GAN benchmark and a substantial performance gain over the CNN-based diffusion model.

The performance advantage of the proposed framework can be attributed to several factors. First, diffusion-based generative modeling provides more stable and expressive distribution learning than adversarial training, enabling more faithful approximation of the underlying THz channel distribution. Second, the transformer-based denoising backbone effectively captures long-range spatial dependencies across the high-dimensional channel representation, whereas CNN-based architectures are inherently constrained by localized receptive fields. This distinction is particularly important for THz UM-MIMO channels, where hybrid near-/far-field propagation effects introduce complex global spatial correlations across the antenna aperture.

Moreover, the angular-domain sparsity of THz channels further facilitates structured generative learning, allowing the proposed model to focus on dominant propagation features rather than dense, noise-like variations. Consequently, the generated channels exhibit substantially improved structural fidelity and spatial consistency compared with the benchmark approaches.

These results validate the effectiveness of the proposed CDiT framework for high-fidelity THz channel generation.

\begin{figure*}[t]
		\centering
		\subfigure[Tx grid.
		]{\includegraphics[width=0.45\textwidth]{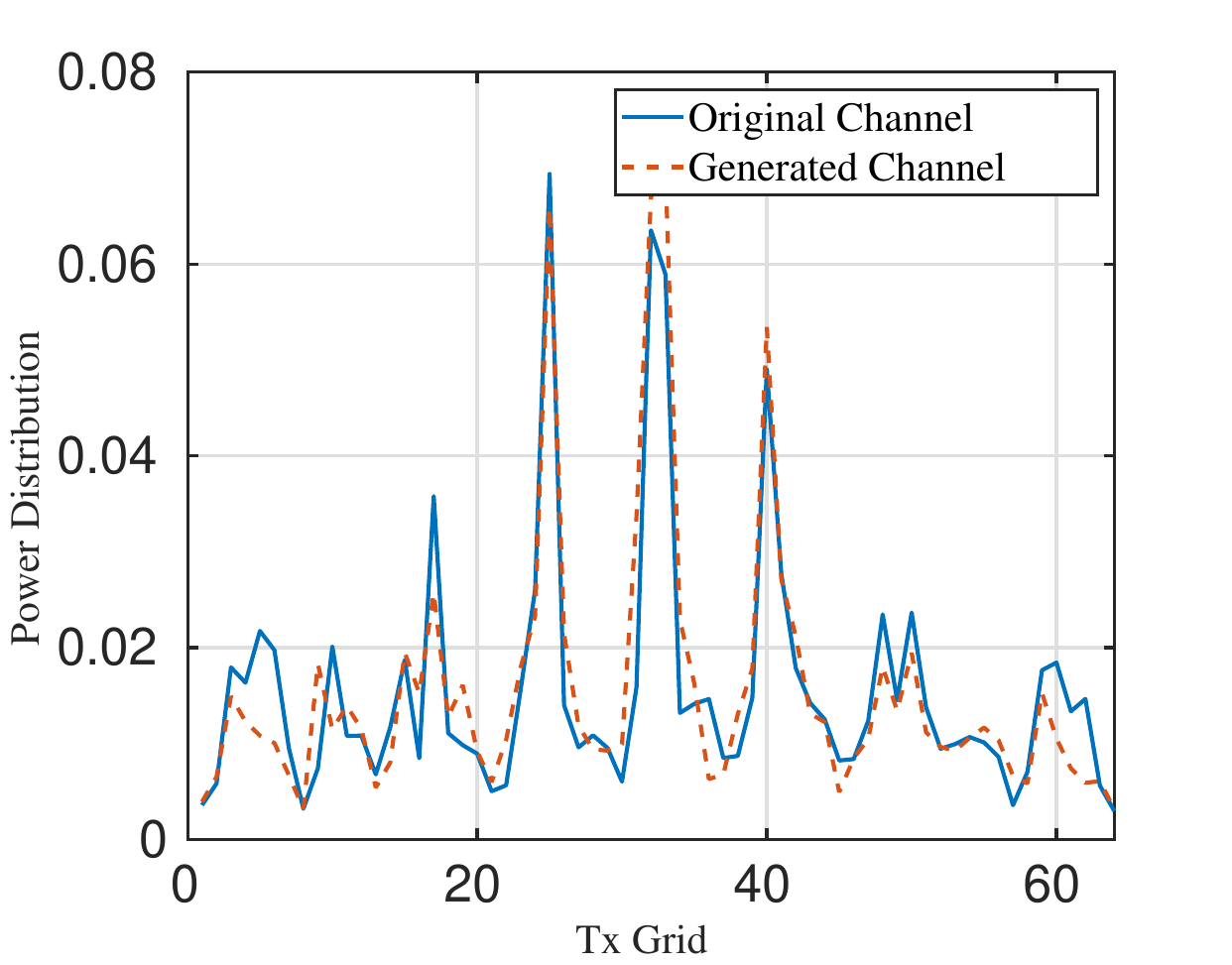}}
		\subfigure[Rx grid.]{\includegraphics[width=0.45\textwidth]{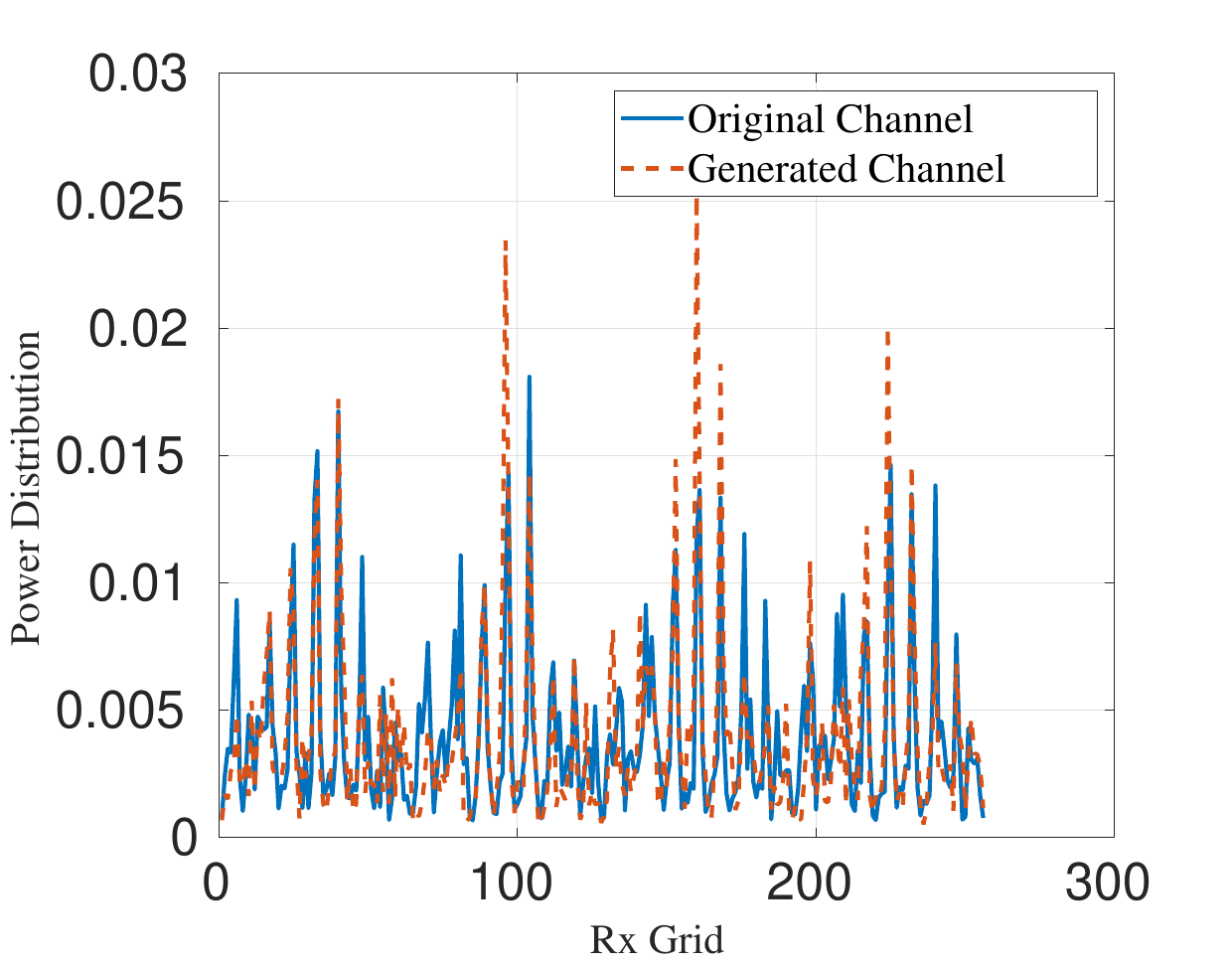}} 
		\caption{Plot of power distribution at Tx grid and Rx grid.}
		\label{fig_power} 
\end{figure*}

\subsection{Power Distribution Evaluation}

After the HPSM transformation, the THz channel is represented in the beamspace domain, where the sparse channel matrix reflects the angular power distribution over the discrete Tx and Rx angular grids induced by the HPSM codebooks. Each grid point corresponds to a quantized propagation direction, providing a physically interpretable representation of the dominant propagation structure and revealing how channel energy is distributed across different spatial directions.

The power distributions of the ground-truth and generated channels are presented in Fig.~\ref{fig_power}. A close agreement can be observed between the generated and reference channels, with the proposed framework accurately preserving the dominant high-energy regions, major propagation clusters, and overall sparse structural patterns in the beamspace domain. In particular, the generated channels successfully reconstruct the principal angular energy concentrations and their corresponding spatial locations, indicating that the model effectively captures the geometry-dependent propagation characteristics of practical THz environments.

Accurate preservation of the beamspace power distribution is particularly important for THz UM-MIMO systems, since beamforming performance strongly depends on reliable characterization of angular energy concentration and dominant propagation directions. Since the beamspace representation directly reflects the spatial distribution of channel energy between the Tx and Rx antenna arrays, consistency in this domain suggests that the essential physical characteristics of the wireless propagation channel are well preserved.

Compared with conventional generative approaches, accurately reconstructing the angular power distribution is a more challenging task, as it requires preserving not only statistical similarity but also physically meaningful spatial structures. The strong consistency observed in Fig.~\ref{fig_power} indicates that the proposed framework does not merely reproduce average channel statistics, but effectively learns the underlying geometry-aware propagation behavior of THz channels.

Together with the structural similarity analysis, these results validate the capability of the proposed framework to achieve high-fidelity THz channel generation in both statistical and physically interpretable spatial domains.

\section{Conclusion}\label{sec:summary}

In this paper, a CDiT framework was proposed for high-fidelity THz channel generation in UM-MIMO systems. By combining the HPSM-based beamspace representation with diffusion modeling and a transformer-based denoising network, the proposed framework enables geometry-aware generation of hybrid near-/far-field THz channels. Experimental results demonstrate stable convergence and clear performance improvements over representative benchmark methods. In particular, the proposed CDiT framework achieves over 70\% improvement in average SSIM compared with the GAN-based baseline and consistently outperforms the CNN-based diffusion model, highlighting its superior capability in preserving complex channel structures. Furthermore, the generated channels closely match the ground-truth channels in terms of both structural similarity and angular power distribution, confirming that the proposed framework effectively captures the essential spatial and statistical characteristics of practical THz propagation environments. These results demonstrate the potential of the proposed approach as a promising data-driven paradigm for THz channel modeling and future wireless system design.

\bibliographystyle{IEEEtran}
\bibliography{main}
\end{document}